# HYPERVIRIAL AND EHRENFEST THEOREMS IN SPHERICAL COORDINATES: SYSTEMATIC APPROACH

ANZOR KHELASHVILI[1,2], TEIMURAZ NADAREISHVILI[1,3]

[1] *Inst. of High Energy Physics, Iv. Javakhishvili Tbilisi State University, University Str. 9, 0109, Tbilisi, Georgia. E-mail:anzor.khelashvili@tsu.ge*

[2] *St.Andrea the First-called Georgian University of Patriarchate of Georgia, Chavchavadze Ave.53a, 0162, Tbilisi, Georgia.*

[3] *Faculty of Exact and Natural Sciences, Iv. Javakhishvili Tbilisi State University, Chavchavadze Ave 3, 0179, Tbilisi, Georgia.*

Corresponding author E-mail: *teimuraz.nadareishvili@tsu.ge*

Abstract: Elaboration of some fundamental relations in 3-dimensional quantum mechanics is considered taking into account the restricted character of areas in radial distance. In such cases the boundary behavior of the radial wave function and singularity of operators at the origin of coordinates contribute to these relations. We derive the relation between the average value of the operator's time derivative and the time derivative of the mean value of this operator, which is usually considered to be the same by definition. The deviation from the known result is deduced and manifested by extra term, which depends on the boundary behavior mentioned above. The general form for this extra term takes place in the hypervirial-like theorems. As a particular case, the virial theorem for Coulomb and oscillator potentials is considered and correction to the Kramers' sum rule is derived. Moreover the corrected Ehrenfest theorem is deduced and its consistency with real physical picture is demonstrated.

Аннотация: В работе рассматривается разработка некоторых фундаментальных отношений в 3-мерной квантовой механике, принимая во внимание ограниченный характер областей в радиальном расстоянии. В таких случаях поведение на границе радиальной волновой функции и особенность операторов в начале координат способствуют этим отношениям. Мы получили отношение между средним значением производной по времени оператора и производной по времени среднего значения этого оператора, который обычно считается тем же по определению. Выведено отклонение от известного результата и это связано появлением дополнительного члена, который зависит от упомянутого выше поведения на границе. Общая форма для этого дополнительного члена проиявляется в гипервириал подобных теоремах. Как частный случай, рассматриваетьтся теорема вириала для Кулоновского и осцилляторного потенциалов и получено поправки к правилам сумм Крамерса. Кроме того, выведена исправленная теорема Еренфеста и продемонстрирована ее соответствие с реальной физической картиной

## 1. Introduction

During decades there have been many studies in which the authors try to reconsider some fundamental relations of quantum mechanics on strong mathematical properties of the operators and their matrix elements appeared. For example, the Ehrenfest theorem or/and, in general, mean values of time derivatives of the operators have been encountered [1-8].

In the textbooks on quantum mechanics most formulations concerned mainly one-dimensional problems, and in these cases, as a rule, wave functions decrease at infinity (Hilbert space). Mostly the problems in full infinite space are considered. However, as is well known, when the system is located in finite volume the inclusion of boundary conditions becomes necessary as well as they impose the restrictions on the allowed classes of wave functions. It is so, because operators are defined not only by their action (i.e, what they do to the function, which they operate on), but also by their domain (that is, the set of functions on which they operate). The situations are encountered frequently, when domain is essential.

This problem often arises in many-dimensional cases, when the polar (spherical) coordinates are necessarily introduced, because the radial functions are defined in semi-space. In such cases problems with restricted area emerge automatically.

Therefore, the question arises: whether or not some of the well-known theorems are altered, when the boundary behavior problem comes into play. The literature on this subject is quite voluminous. Only part of it is listed in References below. Remarkable contributions were made in abovementioned papers, which appeared, by our knowledge, mainly after the pioneering work [1]. Here and in other works strong mathematical definition of operators and their various combinations are established according to their domains. It is also specified how the boundary contributions appear in hypervirial-like relations. As regards of Ehrenfest-like theorems the strong mathematical grounds are derived in [7].

Though this problem was well investigated in one-dimensional case, three dimensions has its specific peculiarities, therefore our attention will be focused to three dimensions. Naturally, some results obtained in one-dimensional cases are general and applicable in three dimensions as well, nevertheless general consideration in three dimensions has its specific interest.

The aim of this article is to study some quantum mechanical theorems with this point of view. We see that in most cases great caution must be exercised especially in the cases, when the potential in the Schrodinger equation is singular or the operators are singular themselves. Our considerations below concern spherically symmetric operators.

This paper is organized as follows: First of all the time derivative of mean value of the operator is studied and the extra term in the usual relation is separated, owing to restricted character of the area in radial space. The obtained surface term depends on the behavior of radial wave function and considered operators in the origin of coordinates. After that new relation between time derivatives is established. This fact has an influence on the relation between mean values and the integrals of motion in various special cases. We demonstrate that the singular character of considered operators has crucial influence on various relations.

The remained part of this manuscript is devoted to some applications of obtained results. Namely, the generalization of hypervirial theorem is considered taking into account this extra term. As a result new form of hypervirial theorem is derived. This modified theorem is verified in case of Coulomb potential and it is shown that the well-known Kramers' theorem should be corrected. We show that the modified version of this theorem works successfully. Lastly the explicit calculations of this extra term for various forms of operators are carried out - The application to the Ehrenfest theorem is considered as well and possible modifications are discussed. It is shown that the obtained extra term plays a role of the so-called "boundary quantum force" and its physical meaning is clarified.

Even though basic ideas, concerning to the Ehrenfest theorem have been discussed and published previously, we did not find a systematic consideration of 3-dimensional problems. We are inclined to think that such cases probably contain many aspects for applications, particularly, for singular operators, which will become clear below.

## 2. Time derivative of the operator's mean value

It is well known that in quantum mechanics derivative of time-dependent operator $\hat{A}(t)$ satisfies the Heisenberg equation

$$\frac{d\hat{A}}{dt} = \frac{\partial \hat{A}}{\partial t} + \frac{i}{\hbar}\left[\hat{H}, \hat{A}\right] \tag{2.1}$$

Averaging this expression by the state functions one derives

$$\left\langle \frac{d\hat{A}}{dt} \right\rangle = \left\langle \frac{\partial \hat{A}}{\partial t} \right\rangle + \frac{i}{\hbar}\left\langle \left[\hat{H}, \hat{A}\right] \right\rangle \tag{2.2}$$

As a rule one believes that these two operations, – time derivative and average procedures, can be interchanged. Let us cite a quotation from the book of Landau and Lifshitz [9] ''The idea of the derivative with respect of time must be differently defined in quantum mechanics. It is natural <u>to define</u> the derivative $\dot{f}$ of a quantity $f$ as a quantity whose mean value is equal to the derivative, with respect to time, of the mean value $\overline{f}$. Thus we have the <u>definition</u> $\overline{\dot{f}} = \dot{\overline{f}}$ ''. (*Underlining is ours*). Therefore, according to this book it is the definition. In several textbooks (see for example [10]), relations (2.1-2.2) are derived from the equations of quantum mechanics, while this problem reduces to definition at long last.

Let us see, if it is valid in general, when the problem is considered in 3-dimensional space. With this aim we calculate

$$\frac{d}{dt}\left\langle \psi \left| \hat{A} \right| \psi \right\rangle = \left\langle \frac{\partial \psi}{\partial t} \middle| \hat{A} \middle| \psi \right\rangle + \left\langle \psi \middle| \frac{\partial \hat{A}}{\partial t} \middle| \psi \right\rangle + \left\langle \psi \middle| \hat{A} \middle| \frac{\partial \psi}{\partial t} \right\rangle \tag{2.3}$$

We use here the time dependent Schrodinger equation and its complex conjugate one

$$i\hbar \frac{\partial \psi}{\partial t} = \hat{H}\psi, \qquad -i\hbar \frac{\partial \psi^*}{\partial t} = (\hat{H}\psi)^* \tag{2.4}$$

Then we have

$$\frac{d\langle \hat{A} \rangle}{dt} = \frac{i}{\hbar}\langle \hat{H}\psi | \hat{A}\psi \rangle - \frac{i}{\hbar}\langle \psi | \hat{A}\hat{H}\psi \rangle + \left\langle \psi \middle| \frac{\partial \hat{A}}{\partial t} \middle| \psi \right\rangle \tag{2.5}$$

Where for any moment of time we must have [4,5]

$$\psi \in Dom(\hat{A}) \cap Dom(\hat{H}) \cap Dom\left(\frac{\partial \hat{A}}{\partial t}\right) \tag{2.6}$$

And

$$\hat{H}\psi \in Dom(\hat{A}) \cap Dom\left(\frac{\partial \hat{A}}{\partial t}\right) \tag{2.7}$$

It must be stressed especially that if the following condition

$$\hat{A}\psi \in Dom(\hat{H}) \tag{2.8}$$

is satisfied [4,5 ], we can introduce a commutator $\left[\hat{H}, \hat{A}\right]$ and rewrite (2.5) in the following way

$$\frac{d\langle \hat{A}\rangle}{dt} = \frac{i}{\hbar}\langle \hat{H}\psi | \hat{A}\psi\rangle - \frac{i}{\hbar}\langle \psi | \hat{H}\hat{A}\psi\rangle + \frac{i}{\hbar}\langle \psi |[\hat{H},\hat{A}]|\psi\rangle + \langle \psi |\frac{\partial \hat{A}}{\partial t}|\psi\rangle \qquad (2.9)$$

The first two terms $\Pi \equiv \frac{i}{\hbar}\langle \hat{H}\psi | \hat{A}\psi\rangle - \frac{i}{\hbar}\langle \psi | \hat{H}\hat{A}\psi\rangle$ are new. They were discovered in [2-6] for one-dimensional case and were calculated there in the simplest models.

One important comment is in order: In writing of Eq. (2.9) the conditions (2.6) –(2.8) are imposed. The constraint (2.8) is the most crucial. When it happens, the additional term vanishes. But, in addition, if the boundary conditions are also imposed, it may be that this restriction fails. For details see [8]. In principle, it is very difficult determine beforehand if this restriction is violated or no. Only detailed calculation can sheds light. We will see below by explicit calculation that the additional term $\Pi$ in (2.9) *does not always disappear*.

Remarkably enough that in [8] the modification of the Heisenberg equation is suggested – foreseeing this result, the authors propose inclusion of the extra terms into the operator equation in advance as follows

$$\frac{dA}{dt} = \frac{\partial A}{\partial t} + \frac{i}{\hbar}[H,A] + \frac{i}{\hbar}(H_* - H)A \qquad (2.10)$$

It is very interesting, but is not always necessary, by our opinion.

Below in contrast with the above-mentioned papers, we consider the 3-dimensional case, when we have arbitrary central potential and $\hat{A}$ operator depends only on radial distance $\hat{A} = \hat{A}(r,...)$. Corresponding radial Hamiltonian is

$$\hat{H} = \frac{1}{2m}\left(-\frac{d^2}{dr^2} - \frac{2}{r}\frac{d}{dr}\right) + \frac{l(l+1)}{2mr^2} + V(r,t) \qquad (2.11)$$

In the process of calculation of additional terms in Eq. (2.9) "redistribution" of radial wave function with Hamiltonian is employed in order to construct the radial Hamiltonian again. For example, the first term in (2.3) looks like

$$I_1 = \langle \frac{\partial \psi}{\partial t}|\hat{A}|\psi\rangle = -\frac{1}{i\hbar}\int_{-\infty}^{\infty}(\hat{H}\psi)^*\hat{A}\psi\, dxdydz = -\frac{1}{i\hbar}\int_0^{\infty}(\hat{H}R)^*\hat{A}Rr^2 dr = -\frac{1}{i\hbar}\int_0^{\infty}\hat{H}R^*\hat{A}Rr^2 dr \qquad (2.12)$$

where $R = R(r)$ is a radial function, $\psi(r) = R(r)Y^{lm}(\theta,\varphi)$. The function $R^*$ needs to be placed at the top of integrand expression. For this replacement only the kinetic part of Hamiltonian operates. Therefore, let us study only the following expression

$$\int_0^\infty \hat{H}R^* \hat{A}Rr^2 dr \Rightarrow \int_0^\infty \left[ -\frac{\hbar^2}{2m}\left(\frac{d^2}{dr^2} + \frac{2}{r}\frac{d}{dr}\right)R^* \right] \hat{A}Rr^2 dr \tag{2.13}$$

Remaining terms of the Hamiltonian do not contribute to the procedure carried out.

Now let us integrate by parts twice in order to transfer differentiation to the right and construct the radial Hamiltonian again. Then we obtain

$$\int_0^\infty \frac{d^2 R^*}{dr^2}\hat{A}Rr^2 dr = \int_0^\infty \hat{A}R \frac{d^2 R^*}{dr^2} r^2 dr = \hat{A}Rr^2 \frac{d^2 R^*}{dr^2}\bigg|_0^\infty - \int_0^\infty \frac{d}{dr}\left(\hat{A}Rr^2\right)\frac{dR^*}{dr} dr =$$
$$= \hat{A}Rr^2 \frac{d^2 R^*}{dr^2}\bigg|_0^\infty - \frac{d}{dr}\left(\hat{A}Rr^2\right)R^*\bigg|_0^\infty + \int_0^\infty R^* \frac{d^2\left(\hat{A}Rr^2\right)}{dr^2} dr \tag{2.14}$$

Here we take into account that for the bound states the radial function tends to zero at partial infinity, but the contribution from the origin, in general, remains. Proceeding this way, we obtains at the end

$$\int \hat{H}\psi^* \hat{A}\psi\, dxdydz = \int \psi^* \hat{H}\hat{A}\psi\, dxdydz + \frac{\hbar^2}{2m}\lim_{r\to 0}\left\{\hat{A}Rr^2 \frac{dR^*}{dr} - r^2 R^* \frac{d}{dr}(\hat{A}R)\right\} \tag{2.15}$$

Considering the same procedure in all the terms, we derive for the required derivative

$$\frac{d\langle \hat{A} \rangle}{dt} = \frac{i}{\hbar}\langle [\hat{H},\hat{A}] \rangle + \left\langle \frac{\partial \hat{A}}{\partial t} \right\rangle + \Pi \tag{2.16}$$

where we obtained for the extra term

$$\Pi = i\frac{\hbar}{2m}\lim_{r\to 0}\left\{ r^2\left[ \hat{A}R \frac{dR^*}{dr} - R^* \frac{d}{dr}(\hat{A}R) \right] \right\} \tag{2.17}$$

It is exactly this term that corresponds to the additional contribution mentioned in Eq. (2.9). This term is not zero in general, because it depends on the behavior of wave function and the operator in the origin of coordinates. Evidently, this term has a purely quantum origin. - It has no classical analogue (in the limit of $\hbar \to 0$, this term also tends to zero). Equation (2.16) together with (2.17) is new. Though analogous relations are shown in Refs. [4-5,8], the derivation in these papers is rather formal. Here, we derived them in explicit form**.**

### 3. Analysis of the additional term

As it is clear from the above, the value of $\Pi$ depends explicitly on the behavior of the radial function at the origin. It is known that under general requirements the radial function must behave like

$$rR(r)\Big|_{r\to 0} = 0 \qquad (3.1)$$

This condition corresponds to the Dirichlet boundary condition for reduced $u = rR$ wave function (for details see [11-14], also the Appendix below). Some authors consider the condition (3.1) as too restrictive and recommends other boundary conditions, which also guarantee the self-adjointness of the reduced radial Hamiltonian. However, in series of articles [11-13] we have shown that the Schrodinger reduced equation is valid only together with Dirichlet boundary condition. For the sake of definiteness we insert the Appendix at the end of this manuscript.

The behavior of reduced wave function, when $r$ turns to the origin of coordinates evidently depends on potential under consideration. From this point of view the following classification is known [15]:

(1). <u>Regular potentials</u>: They behave as

$$\lim_{r\to 0} r^2 V(r) = 0, \qquad (3.2)$$

For which solution at the origin behaves like

$$R\Big|_{r\to 0} = C_1 r^l + C_2 r^{-(l+1)} \qquad (3.3)$$

Clearly, the second term is very singular and contradicts to (3.1). Therefore we must retain only the first term ($C_2 = 0$) or

$$R\Big|_{r\to 0} \approx C_1 r^l \qquad (3.4)$$

- (2). <u>Singular potentials</u>, for which

$$r^2 V(r)\Big|_{r\to 0} \to \pm\infty \qquad (3.5)$$

For them the "falling to the center" happens and is not interesting for us now.

- (3) <u>"Soft" singular potentials</u>, for which

$$r^2 V(r)\Big|_{r\to 0} \to \pm V_0, \quad (V_0 = const > 0) \qquad (3.6)$$

Here the (+) sign corresponds to repulsion, while the (-) sign – to attraction. For such potential the wave function has the following behavior [11-14]:

$$\lim_{r\to 0} R = a_{st} r^{-1/2+P} + a_{add} r^{-1/2-P} \equiv R_{st} + R_{add}, \qquad (3.7)$$

where

$$P = \sqrt{(l+1/2)^2 - \frac{2mV_0}{\hbar^2}} \quad (3.8)$$

In the region $0 < P < 1/2$ the second solution satisfies also the boundary condition (3.1), therefore it must be retained in general and hence the self adjoint extension need to be performed [13]. As for the region $P \geq 1/2$ only the first (standard or regular) solution remains.

Now, let us return to consideration of additional contribution in Eq. (2.16). First, consider regular potentials. It is obvious from Eq. (2.17) that upon calculation of the limit the singularity of the operator $\hat{A}$ in the origin will be also important. We take it as

$$\hat{A}(r) \sim \frac{1}{r^\beta}; \qquad \beta > 0 \quad (3.9)$$

Here, it is implied not only explicit dependence on $r$, but also its scale dimension (derivative et al.). Taking all these into account, we obtain

$$\hat{A} r^l \sim r^{l-\beta} \quad (3.10)$$

Then, we have

$$\Pi_{reg} = \frac{i\hbar C_1^2}{2m} \lim_{r \to 0} r^2 \left\{ r^{l-\beta} l r^{l-1} - r^l \frac{d}{dr} r^{l-\beta} \right\} = \frac{i\hbar C_1^2}{2m} \lim_{r \to 0} r^{2l+1-\beta} \quad (3.11)$$

In order for this expression not to be diverging we must require

$$2l + 1 > \beta \quad (3.12)$$

In this case the additional term vanishes. If the inequality is reflected, then the divergent result will follow and we will be unable to write the equation (2.2).
On the other hand, if the operator is such that

$$2l + 1 = \beta, \quad (3.13)$$

the extra term survives on the right-hand side

$$\frac{d\langle A \rangle}{dt} = \left\langle \frac{\partial \hat{A}}{\partial t} \right\rangle + \frac{i}{\hbar} \langle [\hat{H}, \hat{A}] \rangle + \frac{i\hbar C_1}{m}\left(l + \frac{1}{2}\right) \quad (3.14)$$

We see that the averaging relation is not so trivial, as it looks at the first glance, *but depends on singularity of operator under consideration.*

Let us now return to the case of soft singular potential (3.6). First of all, for the sake of definiteness, ignore the additional contribution ($a_{add} = 0$) and use only the regular solution $R = R_{st} = a_{st} r^{-1/2+P}$:

$$\Pi_{st} = \frac{i\hbar a_{st}^2}{2m} \lim_{r \to 0} r^2 \left\{ r^{-1/2+P-\beta}(-1/2+P) r^{P-3/2} - r^{-1/2+P} \frac{d}{dr} r^{-1/2+P-\beta} \right\} = i\hbar \frac{a_{st}^2 \beta}{2m} \lim_{r \to 0} r^{2P-\beta} \quad (3.15)$$

Here, the index under $\Pi$ indicates that only the regular solutions of (3.6) are considered,
Here also we get $\Pi_{st} = 0$, when $2P > \beta$.

But for

$$2P = \beta \quad (3.16)$$

the finite contribution follows

$$\frac{d\langle \hat{A}\rangle}{dt} = \left\langle \frac{\partial \hat{A}}{\partial t}\right\rangle + \frac{i}{\hbar}\langle [\hat{H},\hat{A}]\rangle + \frac{i\hbar a_{st}^2}{m}P \tag{3.17}$$

According to the obtained results we can conclude that even in ordinary quantum mechanics the well-known averaging relation is valid only in the cases, when the condition (3.12) between orbital momentum and operator singularity at the origin is satisfied. At the same time for "soft" singular potentials the standard solutions must obey the restriction $2P > \beta$. It is evident that this strange result is a consequence of singular character of the considered operator. But it is surprising, that *the time derivative of the average value does not coincide with the average of derivative of the same operator*, if the derivative of the operator is defined by Eq.(2.1). Indeed, from (2.2) and (2.16) it follows

$$\left\langle \frac{d\hat{A}}{dt}\right\rangle = \frac{d\langle \hat{A}\rangle}{dt} - \Pi \tag{3.18}$$

This equation has a principally new meaning. It shows that for singular operators in considered case, (Eq.3.16), the above mentioned *definition* from the classical book [9] is not correct in general. Results of this Section are reflections of imposed conditions (2.6) – (2.8), showing that, the additional terms may at time be present and at other times absent, depending on whether (2.8) is fulfilled or not. (One possible way to keep the balance between derivatives $\bar{\dot{f}} = \dot{\bar{f}}$ is t term beforehand as in (2.10).

Even, when the operator does not depend on time explicitly, the above consideration shows that

$$\frac{d\langle \hat{A}\rangle}{dt} = \frac{i}{\hbar}\langle [\hat{H},\hat{A}]\rangle + \Pi \tag{3.19}$$

Therefore, if the operator has "bad" singularity ((3.13) or (3.16)), its average value is not an integral of the motion, even if it commutes with the Hamiltonian. In the context of this result we think that the meaning of integrals of motion in quantum mechanics must be revised.

In conclusion, we have demonstrated that when one considers the time evolution in spherical coordinates, a definite caution is necessary, in particular, the singular character of the considered operator should be taken into account, as well as the singularity of the wave function itself.

## 4. Stationary states and integrals of motion

Let us now apply the derived results and consider the case when the Hamiltonian doesn't explicitly depend on time. For stationary states wave function has the following dependence

$$\psi(\mathbf{r},t) = e^{-\frac{i}{\hbar}Et}\phi(\mathbf{r}) \tag{4.1}$$

When the operator $\hat{A}$ also doesn't explicitly dependent on time, we should have an operator equality

$$\frac{d\hat{A}}{dt} = \frac{i}{\hbar}\left[\hat{H}, \hat{A}\right] \tag{4.2}$$

Averaging this equality by means of (4.1), we get

$$\left\langle \frac{d\hat{A}}{dt} \right\rangle = \left\langle \frac{i}{\hbar}\left[\hat{H}, \hat{A}\right] \right\rangle, \tag{4.3}$$

Or explicitly

$$\int \psi^* \frac{d\hat{A}}{dt} \psi d^3r = \frac{i}{\hbar} \int \int_0^\infty \phi^* \left[\hat{H}, \hat{A}\right] \phi d^3r = \frac{i}{\hbar} \int_0^\infty R^* \left(\hat{H}\hat{A} - \hat{A}\hat{H}\right) R r^2 dr =$$
$$= \frac{i}{\hbar} \int_0^\infty R^* \hat{H}\hat{A} R r^2 dr - \frac{i}{\hbar} E \int_0^\infty R^* \hat{A} R r^2 dr \tag{4.4}$$

Here, we used the fact, that $\phi$ is an eigenfunction of $\hat{H}$ with eigenvalue $E$. Therefore,

$$\left\langle \frac{d\hat{A}}{dt} \right\rangle = \frac{i}{\hbar} \left\{ \int_0^\infty R^* \hat{H}\hat{A} R r^2 dr - E \int_0^\infty R^* \hat{A} R r^2 dr \right\} \tag{4.5}$$

Let us consider two cases:

(a). $\hat{A}$ commutes with $\hat{H}$. Then it follows

$$\left\langle \frac{d\hat{A}}{dt} \right\rangle = \frac{d\left\langle \hat{A} \right\rangle}{dt} = 0 \tag{4.6}$$

Hence, for stationary state, if the operator $\hat{A}$ is explicitly time-independent and commutes with the Hamiltonian, then in spite of its singular character, the relation (4.6) is valid, that is the mean value of this operator is conserved and is an integral of motion.

(b) Now consider the case when Hamiltonian does not commute with the operator $\hat{A}$. Let us study the following integral entering (4.5)

$$I = \frac{i}{\hbar} \int_0^\infty R^* \hat{H}\hat{A} R r^2 dr \tag{4.7}$$

If we repeat all above consideration again, we derive

$$I = \frac{i}{\hbar} \int_0^\infty \hat{H}R^* \hat{A} R r^2 dr - \Pi = \frac{i}{\hbar} E \int_0^\infty R^* A R r^2 dr - \Pi \tag{4.8}$$

where $\Pi$ is given by Eq. (2.17). Taking this into account, we obtain

$$\left\langle \frac{d\hat{A}}{dt} \right\rangle = -\Pi \tag{4.9}$$

On the other hand, when $\hat{A}$ is independent of time explicitly, we have

$$\langle A \rangle = \int e^{\frac{i}{\hbar}Et} \phi^*(\mathbf{r}) \hat{A} e^{-\frac{i}{\hbar}Et} \phi(\mathbf{r}) d^3\mathbf{r} = \int \phi^*(\mathbf{r}) \hat{A} \phi(\mathbf{r}) d^3\mathbf{r} \tag{4.10}$$

Evidently

$$\frac{d\langle A \rangle}{dt} = 0 \tag{4.11}$$

So we obtain "*strange*" result: for stationary states, in case of non-commutativity $\hat{A}\hat{H} \neq \hat{H}\hat{A}$, Eq. (4.11) is valid or $\langle \hat{A} \rangle$ is conserved, but at the same time according to Eq. (4.9), $\left\langle \frac{d\hat{A}}{dt} \right\rangle \neq 0$. In this particular case this 'strange" result is caused by singularity of operator, $\hat{A}$. Therefore, we conclude from this result that the definition $\overline{\dot{f}} = \dot{\overline{f}}$, given initially, depends on the singularity of the considered operator. *Remark, that this point (operator's singularity) was not discussed in the literature up to now.*

## 5. Modified hypervirial theorems

### *5.1 A general consideration*

Comparing Eqs. (4.3) and (4.9), one derives

$$\frac{i}{\hbar} \left\langle \left[ \hat{H}, \hat{A} \right] \right\rangle = -\Pi \tag{5.1}$$

It follows that the well-known hypervirial theorems should be corrected. The traditional hypervirial theorem is formulated as [16-18]:

*If $\phi$ is a bound state eigenfunction of the Hamiltonian $\hat{H}$ and if $\hat{A}$ is an arbitrary Hermitian time-independent operator involving the coordinates and momenta, then hypervirial theorem for $\hat{A}$ states that*

$$\left\langle \phi, \left[ \hat{H}, \hat{A} \right] \phi \right\rangle = 0 \tag{5.2}$$

It is clear from the above that this theorem must be modified and according to Eq. (5.1) it should have the following form:

$$\langle \phi, [\hat{H}, \hat{A}] \phi \rangle = i\hbar \Pi \quad (5.3)$$

Here the choice of the operator $\hat{A}$ is very important and the definite relations between the average values can be derived. The same form of theorem was previously suggested in [8]

First of all, let us remember the known relations for the Coulomb potential $V = -\dfrac{e^2}{r}$ and oscillator, $V = \dfrac{m}{2}\omega^2 r^2$ as regular potentials [19-21]:

$$2E(s+1)\langle r^s \rangle + e^2(2s+1)\langle r^{s-1} \rangle + \frac{s\hbar^2}{4m}\left[s^2 - (2l+1)^2\right]\langle r^{s-2} \rangle = 0 \quad (5.4)$$

$$2E(s+1)\langle r^s \rangle - m\omega^2(s+2)\langle r^{s+2} \rangle + \frac{s\hbar^2}{4m}\left[s^2 - (2l+1)^2\right]\langle r^{s-2} \rangle = 0 \quad (5.5)$$

It is noted in textbooks and various articles, that these relations are valid only if $s > -(2l+1)$. There are papers [22-23], in which these relations are modified for arbitrary N-dimensional Schrodinger equation to have a form

$$(2L+1)^2 C_l^2 \delta_{S,-2L} = \frac{2m}{\hbar^2}\left\{\left\langle r^S \frac{dV}{dr}\right\rangle + 2S\langle r^{S-1}V\rangle - 2SE\langle r^{S-1}\rangle\right\} + \frac{1}{2}(S-1)\left[(2L+1)^2 - (S-1)^2\right]\langle r^{S-3}\rangle \quad (5.6)$$

Where $L = l + \dfrac{N-3}{2}$ and $\lim_{r \to 0} r^{-l} R_l(r) = C_l$

Note that the relation (5.6) was earlier derived in [24] by different method for 3-dimensional case.

In our paper [25] significantly more general relations were derived. Namely, we considered the general second order differential equation

$$R''(r) + \frac{2}{r}R'(r) + L(r)R(r) = 0 \quad (5.7)$$

This equation reduces to the known equations (radial Schrodinger, one- and two- body Klein-Gordon etc.). Then, after multiplication of Eq. (5.7) on an arbitrary three-times differentiable function $f(r)$ and partial integration we derived very general hypervirial theorem (see, [25])

$$\left\{ f\left[R^2 - r^2 RR'' + r^2 R'^2\right] - f'rR[rR' + R] + \frac{1}{2}f''r^2 R^2 \right\}_{r=0} = -2<fL> - <fL'> - \frac{1}{2}<f'''> \quad (5.8)$$

from which by choosing $f(r)$, one can obtain several interesting relations. Some of them are exhibited in mentioned paper.

### 5.2 Application of the modified hypervirial theorem

Consider now some applications of a new modified hypervirial theorem (5.3). We'll see below that the expression of $\Pi$ derived above is not sufficient for all cases. For some operators $A$ it becomes necessary to go into details, depending on its singular character. First of all, let analyze cases which were discussed earlier in literature. Consider the following operator [19]

$$A = \hat{p}_r f(r) \quad (5.9)$$

where $\hat{p}_r$ is a radial momentum (hermitian) operator [9-10]

$$\hat{p}_r = \frac{\hbar}{i}\left(\frac{\partial}{\partial r} + \frac{1}{r}\right) \quad (5.10)$$

and $f(r)$ is a three-times differentiable. . Calculate the commutator

$$[\hat{H}, \hat{A}] = -\frac{i\hbar}{2m}\left\{\hat{p}_r^2 \frac{df}{dr} + \frac{df}{dr}\hat{p}_r^2\right\} - \frac{\hbar^2}{2m}\frac{d^2 f}{dr^2}\hat{p}_r + i\hbar f(r)\frac{dV}{dr} - i\hbar^3 \frac{l(l+1)}{mr^3}f \quad (5.11)$$

Entering here $\hat{p}_r^2$ and $\hat{p}_r$ rewrite as

$$-\frac{\hat{p}_r^2}{2m} = -H + \frac{\hbar^2 l(l+1)}{2mr^2} + V \quad (5.12)$$

$$-\frac{\hbar}{m}i\frac{d^2 f}{dr^2}\hat{p}_r = \left[H, \frac{df}{dr}\right] + \frac{\hbar^2}{2m}f''' \quad (5.13)$$

Finally

$$[\hat{H}, \hat{A}] = i\hbar Q - \frac{3}{2}i\hbar[\hat{H}, f'] \quad (5.14)$$

where

$$Q = -2f'(H-V) + \hbar^2 \frac{l(l+1)}{m}\left[\frac{f'}{r^2} - \frac{f}{r^3}\right] - \frac{\hbar^2}{4m}f''' + f(r)V' \quad (5.15)$$

Then from (5.3), (5.14) and (5.15) it follows

$$\langle Q \rangle - \frac{3}{2}\langle [\hat{H}, f'] \rangle = \Pi \tag{5.16}$$

Here $\Pi$ is given by (2.17).

This place is principally important. In [26-27] analogous relations was studied for the following operator

$$\hat{A}_1 = f(r)\hat{p}_r \tag{5.17}$$

It differs from the above operator (5.9) by permutation

$$\hat{A}_1 = \hat{A} + i\hbar \frac{df}{dr} \tag{5.18}$$

and there was remarked that –"the hypervirial theorem demands the expectation values of both $[H, A_1]$ and $[H, f']$ to be zero".

But it is not so. In particular, *the expectation value of $[H, f']$ is not zero*. Indeed, if we use the relation (2.16) for $\frac{d\langle f' \rangle}{dt}$ or repeat a direct calculation for this average, we obtain

$$\frac{d\langle f' \rangle}{dt} = \frac{i}{\hbar}\langle [\hat{H}, f'] \rangle + \Pi' \tag{5.19}$$

Where

$$\Pi' = i\frac{\hbar}{2m}\lim_{r \to 0}\left\{ r^2 \left[ fR\frac{dR^*}{dr} - R^*\frac{d}{dr}(fR) \right] \right\} = -i\frac{\hbar}{2m}\lim_{r \to 0}\{r^2 R^2 f''\} \tag{5.20}$$

From these relations it follows ones again that the singularity of $\hat{A}$ operator ($f$ in this case) participates into calculations. For stationary states $\frac{d\langle f' \rangle}{dt} = 0$, therefore (5.19) reads

$$\langle [\hat{H}, f'] \rangle = i\hbar\Pi' \tag{5.21}$$

It follows from (5.20) and (5.21) that

$$\langle [\hat{H}, f'] \rangle = \frac{\hbar^2}{2m}\lim_{r \to 0}\{r^2 R^2 f''\} \tag{5.22}$$

and from (5.16) and (5.22) that

$$\langle Q \rangle = \Pi + \frac{3\hbar^2}{4m}\lim_{r \to 0}\{r^2 R^2 f''\} \tag{5.23}$$

Now by using (2.17), (5.9),(5.10) let us calculate the following expression

$$\Pi_{tot} = \Pi + \Pi' = \frac{\hbar^2}{2m} \lim_{r \to 0} \left\{ r^2 \left[ \left( \frac{1}{\partial r} + \frac{1}{r} \right) fR \frac{dR^*}{dr} - R^* \frac{d}{dr} \left( \frac{1}{\partial r} + \frac{1}{r} \right) fR \right] \right\} =$$
$$= \frac{\hbar^2}{2m} \lim_{r \to 0} \left\{ r^2 \left[ -R^2 f'' - \left( RR' + \frac{R^2}{r} \right) f' + \left( R'^2 - RR'' + \frac{R^2}{r^2} \right) f \right] \right\}$$

(5.24)

Taking into account (5.24), (5.23) and (5.15) we obtain the most general hypervirial theorem for the Schrodinger equation

$$\left\langle -2f'(H-V) + \hbar^2 \frac{l(l+1)}{m} \left[ \frac{f'}{r^2} - \frac{f}{r^3} \right] - \frac{\hbar^2}{4m} f''' + f(r)V' \right\rangle =$$
$$= \frac{\hbar^2}{2m} \lim_{r \to 0} \left\{ r^2 \left[ -R^2 f'' - \left( RR' + \frac{R^2}{r} \right) f' + \left( R'^2 - RR'' + \frac{R^2}{r^2} \right) f + \frac{3}{2} R^2 f'' \right] \right\}$$

(5.25)

This equation coincides with above mentioned general equation (5.8), when for the operator $L$ we take

$$L = \frac{2m}{\hbar^2} \left[ E - V - \frac{\hbar^2 l(l+1)}{2mr^2} \right]$$

(5.26)

The analogous relation was derived in [28], but it is applicable only for the reduced Schrodinger equation, i.e. in case of regular potentials as we shown in [11-13].

Contrary to that our above derived relation (5.25) can be used in arbitrary case. Let consider some examples.

(a). For the standard solution of "soft" singular potential (3.6) we have

$$\Pi_{tot,sing} = \frac{\hbar^2 a_{st}^2}{4m} \lim_{r \to 0} \left[ f'' r^{2P+1} + (2P+1)\left( \frac{f}{r} - f' \right) r^{2P} \right]$$

(5.27)

Therefore our theorem (5.25) reads

$$\left\langle -2f'(H-V) + \hbar^2 \frac{l(l+1)}{m} \left[ \frac{f'}{r^2} - \frac{f}{r^3} \right] - \frac{\hbar^2}{4m} f''' + f(r)V' \right\rangle =$$
$$= \frac{\hbar^2 a_{st}^2}{4m} \lim_{r \to 0} \left[ f'' r^{2P+1} + (2P+1)\left( \frac{f}{r} - f' \right) r^{2P} \right]$$

(5.28)

Now, if

$$\lim_{r \to 0} fr^{2P-1} = 0 , \qquad (5.29)$$

it follows

$$\Pi_{tot,sing} = 0 \qquad (5.30)$$

Moreover, if

$$\lim_{r \to 0} fr^{2P-1} = const \qquad (5.31)$$

then

$$\Pi_{tot,\sin gular} = \frac{2\hbar^2 a_{st}^2}{m} P^2 \qquad (5.32)$$

It follows from restrictions (5.29), (5.31) that, for example, for regular potentials, there appear some "critical" singular (5.9) like operators, for which $\Pi$ is done by above mentioned relations. For example,

$$\text{For } \begin{cases} l = 0,\ \hat{A} = \hat{p}_r \\ l = 1,\ \hat{A} = \hat{p}_r \dfrac{1}{r^2} \quad \text{etc.} \\ l = 2,\ \hat{A} = \hat{p}_r \dfrac{1}{r^4} \end{cases} \qquad (5.33)$$

If in (5.28) we take a particular case, considered in [18]

$$f = r^{s+1} \qquad (5.34)$$

we obtain

$$\frac{4m}{\hbar^2}\left\{\left\langle r^{s+1}\frac{dV}{dr}\right\rangle + 2(s+1)\left[\left\langle r^s V\right\rangle - E\left\langle r^s\right\rangle\right]\right\} + s\left[(2l+1)^2 - s^2\right]\left\langle r^{s-2}\right\rangle = a_{st}^2 s(s-2P)\delta_{2P,-s} \qquad (5.35)$$

(b). For the regular potential (3.2) $P = l + \dfrac{1}{2}$ and we have

$$\frac{2m}{\hbar^2}\left\{\left\langle r^{s+1}\frac{dV}{dr}\right\rangle + 2(s+1)\left[\left\langle r^s V\right\rangle - E\left\langle r^s\right\rangle\right]\right\} + \frac{1}{2}s\left[(2l+1)^2 - s^2\right]\left\langle r^{s-2}\right\rangle = (2l+1)^2 C_l^2 \delta_{s+1,-2l} \qquad (5.36)$$

This form coincides with eq. (5.6) if in Eq. (5.36) we replace $s+1 \to s$, which means that calculation by commutator gives the same result as a calculation by means of integration by part. In conclusion we can say that the modified hypervirial theorems for the Coulomb and oscillator potentials have the following forms, correspondingly

1. $2E(s+1)\langle r^s \rangle + e^2(2s+1)\langle r^{s-1} \rangle + \dfrac{s\hbar^2}{4m}\left[s^2 - (2l+1)^2\right]\langle r^{s-2} \rangle = -\dfrac{\hbar^2}{2m}(2l+1)^2 C_l^2 \delta_{S+1,-2l}$  (5.37)

2. $2E(s+1)\langle r^s \rangle - m\omega^2(s+2)\langle r^{s+2} \rangle + \dfrac{s\hbar^2}{4m}\left[s^2 - (2l+1)^2\right]\langle r^{s-2} \rangle = -\dfrac{\hbar^2}{2m}(2l+1)^2 C_l^2 \delta_{S+1,-2l}$  (5.38)

We see that the difference of these relations from those of (5.4)-(5.5) consists in the right-hand sides of the given forms. Exactly these sides balance obtained sum rules, discussed below.

Let us make *two comments*:

1. For $s=0$ from (5.37-38) follows the *usual* virial theorem. So in this case the usual virial theorem is correct.
2. For $s=-1$ or $f = const$ it follows from (5.20) that $\Pi' = 0$. This case will be considered below in connection with Ehrenfest theorem.

For the verification of derived results the known solvable potential models are considered more frequently in the current literature. Therefore below we check validity of above sum rules (5.37) and (5.38) for the Coulomb and oscillator potentials. Moreover we include here other interesting operators.

## 6. The cases of Coulomb and oscillator potentials

**(a)** *Coulomb Potential*

Consider for more details the Coulomb potential $V = -\dfrac{e^2}{r}$. Its wave function is [9]

$$R_{nl}(r) = \tilde{C}_{nl}\left(\dfrac{B}{n}\right)^l r^l e^{-\frac{Br}{2n}} F\left(-n+l+1, 2l+2, \dfrac{Br}{n}\right),$$  (6.1)

where

$$\tilde{C}_{nl} = \dfrac{B}{n^2(2l+1)!}\left(\dfrac{B}{2}\right)^{\frac{1}{2}} \sqrt{\dfrac{(n+l)!}{(n-l-1)!}}$$  (6.2)

is a normalization constant, which is related to $C_l$ as follows

$$C_l = \tilde{C}_{nl}\left(\dfrac{B}{n}\right)^l$$  (6.3)

and

$$\frac{B}{n} = \frac{2}{na_0}, \qquad a_0 = \frac{\hbar^2}{me^2} \tag{6.4}$$

where $a_0$ is a Bohr's first orbit radius. Substituting all this into Eq. (5.37), we derive a modified Kramers' relation

$$2E(s+1)\langle r^s \rangle + e^2(2s+1)\langle r^{s-1} \rangle + \frac{s\hbar^2}{4m}\left[s^2 - (2l+1)^2\right]\langle r^{s-2} \rangle = -\frac{\hbar^2}{2m}(2l+1)^2 \widetilde{C}_{nl}^2 \left(\frac{2}{na_0}\right)^{2l} \delta_{s+1,-2l} \tag{6.5}$$

Let us study this relation. It is clear that when $s = -(2l+1)$, then it follows

$$4El\langle r^{-2l-1} \rangle + e^2(4l+1)\langle r^{-2l-2} \rangle = \frac{\hbar^2}{2m}(2l+1)^2 \widetilde{C}_{nl}^2 \left(\frac{2}{na_0}\right)^{2l} \tag{6.6}$$

For verification of its validity, consider some of first values of $l$:

(i) $l = 0$.

This case corresponds to $s = -1$, i.e. $\hat{A} = \hat{p}_r$. Then Eq. (6.5) gives

$$e^2 \left\langle \frac{1}{r^2} \right\rangle = \frac{\hbar^2}{2m} C_{n0}^2 \tag{6.7}$$

It means that if we take zero on the right-hand side (or use the Kramers' relation (5.3)) we'll get the obvious contradiction - $\langle r^{-2} \rangle = 0$. It must be pointed out that the above considered case lies outside the validity of Kramers' relation. *So our theorem generalizes the Kramers' relation.*

Now let us check if the formula (6.6) is fulfilled. The matrix elements of some degrees of radius for the Coulomb functions are known. For instance [9]

$$\left\langle \frac{1}{r^2} \right\rangle = \frac{2}{a_0^2 n^3 (2l+1)} \tag{6.8}$$

For the case under consideration we have - $C_0 = \widetilde{C}_{n0}$,

$$\widetilde{C}_{n0} = \frac{B}{n^2}\left(\frac{B}{2}\right)^{\frac{1}{2}} \sqrt{\frac{n!}{(n-1)!}} = \frac{B}{n^2}\left(\frac{B}{2}\right)^{\frac{1}{2}} \sqrt{n} \quad \text{or} \quad \widetilde{C}_{n0}^2 = \frac{B^2}{n^3}\frac{B}{2}. \tag{6.9}$$

But $\dfrac{B}{2} = \dfrac{1}{a_0}$, and therefore

$$C_0^2 = \widetilde{C}_{n0}^2 = \frac{4}{n^3 a_0^3} \tag{6.10}$$

After substitution all of these into (6.7), we obtain the identity

$$\frac{\hbar^2}{2m}\frac{4}{n^3 a_0^3} = \frac{2e^2}{n^3 a_0^2}, \tag{6.11}$$

from which a correct relation for the Bohr's radius follows.

Hence, *the modified Kramers' relation is successful.*

(ii) $l = 1$

In this case $s = -3$ and corresponding operator is

$$\hat{A} = \hat{p}_r \cdot \frac{1}{r^3} \qquad (6.12)$$

Then Eq. (6.6) gives

$$4E\langle r^{-3}\rangle + 5e^2 \langle r^{-4}\rangle = \frac{\hbar^2}{2m} 9\widetilde{C}_{nl}^2 \left(\frac{2}{na_0}\right)^2 \qquad (6.13)$$

Using here known relations [9]

$$\left\langle \frac{1}{r^3} \right\rangle = \frac{2}{a_0^3 n^3 l(l+1)(2l+1)}, \qquad \left\langle \frac{1}{r^4} \right\rangle = \frac{[3n^2 - l(l+1)]}{2a_0^4 n^5 (l+3/2)(l+1)(l+1/2)l(l-1/2)}. \qquad (6.14)$$

It is easy to verify that the relation (6.6) is also satisfied precisely. So are for $l = 2$ and etc.

### (b) The oscillator potential

Consider now the oscillator potential

$$V = \frac{m}{2}\omega^2 r^2 \qquad (6.15)$$

The wave functions for it are [9]

$$R_{n,l}(r) = \sqrt{2}\alpha^{\frac{2l+3}{4}} \left[\frac{(l+3/2)(l+5/2)\ldots(l+n_r+1/2)}{\Gamma(l+3/2)n_r!}\right]^{1/2} r^l e^{-\frac{\alpha r^2}{2}} F(-n_r, l+3/2, \alpha r^2); \qquad (6.16)$$

for $n_r = 1, 2, 3, \ldots$

and

$$R_{0l} = \sqrt{2}\alpha^{\frac{2l+3}{4}} [\Gamma(l+3/2)]^{-1/2} r^l e^{-\frac{\alpha r^2}{2}}; \quad \alpha = \frac{m\omega}{\hbar}, \quad \text{for } n_r = 0, \qquad (6.17)$$

Comparison with (3.3) gives

$$C_{n,l} = \sqrt{2}\alpha^{\frac{2l+3}{4}} \left[\frac{(l+3/2)(l+5/2)\ldots(l+n_r+1/2)}{\Gamma(l+3/2)n_r!}\right]^{1/2}; \qquad n_r = 1, 2, \ldots \qquad (6.18)$$

and

$$C_{0l} = \sqrt{2}\alpha^{\frac{2l+3}{4}}\left[\Gamma(l+3/2)\right]^{-1/2}; \qquad n_r = 0 \qquad (6.19)$$

Substituting all of this into Eq. (5.38) the modified Kramers' relations take the form

$$2E(s+1)\langle r^s \rangle - m\omega^2(s+2)\langle r^{s+2} \rangle + \frac{s\hbar^2}{4m}\left[s^2 - (2l+1)^2\right]\langle r^{s-2} \rangle =$$
$$= -\frac{\hbar^2}{2m}(2l+1)^2 C_{n_r,l}^2 \delta_{s+1,-2l}; \qquad n_r = 0,1,2.. \qquad (6.20)$$

It is clear that for $s = -(2l+1)$ it follows

$$4El\langle r^{-2l-1} \rangle + m\omega^2(1-2l)\langle r^{1-2l} \rangle = \frac{\hbar^2}{2m}(2l+1)^2 C_{n_r,l}^2; \; n_r = 0,1,2... \qquad (6.21)$$

Consider some of first values of $l$.

i) $l = 0$

This case corresponds to $S = -1$, i.e. $\hat{A} = \hat{p}_r$, one obtains from (6.21)

$$m\omega^2 \langle r \rangle = \frac{\hbar^2}{2m} C_{n_r,0}^2; \qquad n_r = 0,1,2.. \qquad (6.22)$$

It shows that if we made use the usual Kramers' relation (5.5), we get the obvious contradiction, $\langle r \rangle = 0$. It can be noted that in this case from (5.36) we have a general conclusion

$$\frac{C_0^2 \hbar^2}{2m} = \left\langle \frac{dV}{dr} \right\rangle \qquad (6.23)$$

Moreover, from (3.3) follows

$$C_0 = R_0(0) = \sqrt{4\pi}\psi_{00}(0) \qquad (6.24)$$

where

$$\psi_{00} = \frac{R_0}{\sqrt{4\pi}} \qquad (6.25)$$

is a full wave function. Therefore we have derived the well-known relation [29]

$$|\psi_{00}(0)|^2 = \frac{m}{2\pi\hbar^2}\left\langle \frac{dV}{dr} \right\rangle \qquad (6.26)$$

Thus inclusion of $\Pi$ term provides the correct results. Without it (Kramers' case) this was not possible.

Now let us verify correctness of the mean radius relation (6.21) for $n_r = 0$. It is easy to show, that

$$\langle r \rangle = C_{00}^2 \frac{1}{2\alpha^2} \tag{6.27}$$

And if we insert this relation into (6.21), it follows

$$m\omega^2 C_{00}^2 \frac{1}{2\alpha^2} = \frac{\hbar^2}{2m} C_{00}^2 \tag{6.28}$$

which becomes an identity as well as $\alpha = \dfrac{m\omega}{\hbar}$. So this case gives correct result.

Now let us investigate more general problem, $n_r = 1,2,3...$ In this case we can use the following integrals from the Appendix of [9]

$$\int_0^\infty e^{-kz} z^{\nu-1} [F(-n,\gamma,kz)]^2 dz = \frac{\Gamma(\nu)n!}{k^\nu \gamma(\gamma+1)...(\gamma+n-1)} \times$$
$$\times \left[ 1 + \sum_{s=0}^{n-1} \frac{n(n-1)...(n-s)(\gamma-\nu-s-1)(\gamma-\nu-s)...(\gamma-\nu+s)}{[(s+1)!]^2 \gamma(\gamma+1)...(\gamma+s)} \right] \tag{6.29}$$

Making use of this form in general is rather tremendous. Therefore without the loss of generality we consider only $n_r = 1$ case. Now for any $l$ we derive

$$\langle r \rangle = C_{1l}^2 \frac{1}{8\alpha^{l+2}} \frac{(l+1)!}{(l+3/2)^2} (4l+9) \tag{6.30}$$

Using (6.18) for $l = 0$, it follows

$$\langle r \rangle = C_{10}^2 \frac{1}{2\alpha^2} \tag{6.31}$$

And after inserting of (6.31) into (6.21) we check that (6.21) is also correct.

(ii)    $l = 1$

In this case $S = -3$ and corresponding operator is

$$\hat{A} = \hat{p}_r \cdot \frac{1}{r^3} \tag{6.32}$$

Then Eq. (6.21) gives

$$4E\langle r^{-3}\rangle - m\omega^2 \langle r^{-1}\rangle = \frac{\hbar^2}{2m} 9 C^2_{n_r 1}; \qquad n_r = 0,1,2... \qquad (6.33)$$

Consider again two cases:

(a) $n_r = 0$

Now Eq.(6.29) gives

$$\left\langle \frac{1}{r} \right\rangle = \frac{C^2_{01}}{2\alpha^2}; \qquad \left\langle \frac{1}{r^3} \right\rangle = \frac{C^2_{01}}{\alpha} \qquad (6.34)$$

By using $E_{01} = \frac{5}{2}\hbar\omega$ and $\alpha = \frac{m\omega}{\hbar}$ it is easy exercise to convince that (6.33) is valid

(b) $n_r = 1,2,..$

Now the Eq. (6.29) gives in case $n_r = 1; l = 1$

$$\left\langle \frac{1}{r} \right\rangle = \frac{C^2_{11}}{\alpha^2} \frac{9}{50}; \qquad \left\langle \frac{1}{r^3} \right\rangle = \frac{C^2_{11}}{\alpha^2} \frac{13}{50} \qquad (6.35)$$

Inserting this into (6.33) and using $E_{11} = \hbar\omega \frac{9}{2}$ and $\alpha = \frac{m\omega}{\hbar}$ we prove the identity, or (6.33) is correct.

So is for $l = 2$ and etc.

# 7. Modification of the Ehrenfest theorem

As is well-known, the Ehrenfest's equations signify that the average values of position and linear momentum operators evolve classically. The heuristic justification can be found in any quantum mechanical textbooks. However, a rigorous version of this theorem under satisfactory assumptions with standard functional analytic arguments was pointed out in [7], (See, also [30]).

We do not have a claim on such stronger discussion, but simply analyze what happens with the Ehrenfest theorem in ordinary quantum mechanics in light of the influence of presented boundary behavior in spherical coordinates.

Consider again the operator of radial momentum

$$\hat{A} = \hat{p}_r = \frac{\hbar}{i}\left(\frac{\partial}{\partial r} + \frac{1}{r}\right) \qquad (7.1)$$

Substitute it into Eq. (3.19), we have

$$\frac{d\langle \hat{p}_r \rangle}{dt} = \frac{i}{\hbar}\langle [\hat{H}, \hat{p}_r] \rangle + \Pi_{st}, \qquad (7.2)$$

where

$$\Pi_{st} = \frac{a_{st}^2 \hbar^2}{2m}\lim_{r\to 0} r^2 \left\{ \left(\frac{\partial}{\partial r} + \frac{1}{r}\right)[r^{-1/2+P}]\left(-\frac{1}{2}+P\right)(r^{-3/2+P}) - (r^{-1/2+P})\frac{d}{dr}A\left(\frac{\partial}{\partial r} + \frac{1}{r}\right)[r^{-1/2+P}] \right\} = \\ = \frac{a_{st}^2 \hbar^2}{2m}\left(\frac{1}{2}+P\right)\lim_{r\to 0} r^{2P-1} \qquad (7.3)$$

It is clear from this relation that $\Pi_{st} = 0$ for $2P > 1$, while for $2P < 1$, it diverges. But for $2P = 1$ it survives

$$\Pi_{st} = \frac{a_{st}^2 \hbar^2}{2m} \qquad (7.4)$$

Therefore, for singular potential the usual Ehrenfest theorem

$$\frac{d\langle \hat{p}_r \rangle}{dt} = \frac{i}{\hbar}\langle [\hat{H}, \hat{p}_r] \rangle \qquad (7.5)$$

is applicable only in the first case, $2P > 1$. In other cases the additional term (7.4) appears or theorem has no place at all. Remember that *in traditional textbooks this fact is not mentioned.*

Let us now calculate the commutator in (7.5). We find

$$[\hat{H}, \hat{p}_r] = \frac{\hbar^2 l(l+1)}{2m}\left[\frac{1}{r^2}, p_r\right] + [V(r), p_r] \qquad (7.6)$$

But

$$\left[\frac{1}{r^2}, p_r\right] = -i\frac{2\hbar}{r^3}; \qquad [V, p_r] = i\hbar\frac{\partial V}{\partial r} = -i\hbar F_r \qquad (7.7)$$

Where $F_r$ is a "radial force". Therefore we get

$$[\hat{H}, \hat{p}_r] = -i\frac{\hbar^3 l(l+1)}{mr^3} - i\hbar F_r \qquad (7.8)$$

And after taking into account an additional contribution (7.2) we obtain the modified Ehrenfest theorem for time evolution of radial momentum (Newton's "second law):

$$\frac{d\langle \hat{p}_r \rangle}{dt} = \frac{\hbar^2 l(l+1)}{m}\left\langle \frac{1}{r^3} \right\rangle + \langle F_r \rangle + \Pi_{st} \tag{7.9}$$

*This relation is a new one also.* which is a "master equation" and its physical meaning is elucidated for the Coulomb potential once again -  It is remarkable to note that in [2,3] the Ehrenfest theorem in one-dimensional Schrodinger equation was considered in finite interval $(0, a)$ and in semi-axis $(-\infty, 0)$. They derived a formula

$$\frac{d}{dt}\langle \hat{p} \rangle = \langle f_B \rangle \tag{7.10}$$

The authors pointed out that $f_B$ can be considered as a boundary quantum force. Note that for a particle in an infinite square well potential the boundary term (7.10) is zero. In fact, in the open interval $\Omega = (-\infty, \infty)$ the solution and its derivative through x tends to zero when $x \to \pm\infty$. However, in this case the mean value of the external classical force does not vanish [2]

By comparing (7.10) with (7.9), and light of the fact that in spherical coordinates radial variable changes in semi-axis $(0, \infty)$, one can identify $\Pi_{st}$ with the boundary quantum force $f_B$. It is evident that if we turn to the one-dimensional case: $r \to x \in (-\infty, \infty)$ and take $l = 0$, then according of above discussion we derive $\Pi = 0$ and  from (7.9) the true one-dimensional Ehrenfest theorem follows $\frac{d\langle p_x \rangle}{dt} = \langle F_x \rangle$

For regular potentials, when $P = l + 1/2$, only in case $l > 0$ follows $\Pi_{reg} = 0$. As for $l = 0$ it follows

$$\Pi_{reg} = \frac{C_1^2 \hbar^2}{2m}, \qquad \text{evidently} \qquad C_1 = C_0 \tag{7.11}$$

We conclude here that *for regular potentials the usual Ehrenfest theorem is valid only in case $l > 0$, but in case -  $l = 0$ there appears an extra term (7.11).*

Now let us show that Eq. (7.9) gives correct results for Coulomb potential. First consider the case -  $l > 0$. In this case $\Pi_{reg} = 0$. In [31- 32] right-hand side of theorem consists only real forces $\frac{\hbar^2 l(l+1)}{m}\left\langle \frac{1}{r^3} \right\rangle + \langle F_r \rangle$: the sum of radial and centrifugal forces.  In the hydrogen atom problem these two forces compensate each other. Indeed, $F_r = -\frac{e^2}{r^2}$ and using known matrix elements for Coulomb functions

$$\left\langle \frac{1}{r^2} \right\rangle = \frac{1}{n^3 a_0^2}\frac{1}{(l+1/2)}, \qquad \left\langle \frac{1}{r^3} \right\rangle = \frac{1}{n^3 a_0^3 l(l+1)(l+1/2)} \tag{7.12}$$

It is an easy exercise to convince that these two forces compensate each other exactly. So the Newton's second law is satisfied.

On the other hand, the case $l = 0$ is more interesting and crucial. In this case we have no centrifugal term, and the additional term is given by (7.11),

$$\frac{d\langle p_r \rangle}{dt} = \langle F_r \rangle + \frac{C_1^2 \hbar^2}{2m} \tag{7.13}$$

At the same time (see, Eq. (7.12))

$$\langle F_r \rangle = -\frac{2e^2}{n^3 a_0^2} \tag{7.14}$$

As it was mentioned above, in stationary case the left-hand side of (7.13) must be zero. So we should have

$$\frac{C_1^2 \hbar^2}{2m} = \frac{2e^2}{n^3 a_0^2} \tag{7.15}$$

and according to eq. (6.10), it follows a correct expression for Bohr's first orbit radius, $a_0 = \frac{\hbar^2}{me^2}$

It is evident that without the extra term we should have

$$\frac{d\langle p_r \rangle}{dt} = \langle F_r \rangle \tag{7.16}$$

which is clear contradiction (!)

We conclude that in Eq. (7.9) the term $\Pi_{st}$ is necessary for deriving correct results, which is absent in [31-32]. So we have shown that for the $l = 0$ state our result differs from that, which is known in current literature – the source of difference lies in relations (7.13-16). As it is obvious from the definitions (7.4) and (7.11) both $\Pi_{st} > 0$ and $\Pi_{reg} > 0$, so in both cases the quantum boundary force is repulsive, "so it causes the "center of mass" of quantum packet to move far from the boundary" [33].

Lastly, consider the Ehrenfest theorem for the coordinate operator, $\hat{A} = \hat{r}$. Inserting this operator into the definition (2.17), we find

$$\Pi = i\frac{a_{st}^2 \hbar^2}{2m} \lim_{r \to 0} r^2 \left\{ r\left[r^{-1/2+P}\right]\left(-\frac{1}{2}+P\right)\left(r^{-3/2+P}\right) - \left(r^{-1/2+P}\right)\frac{d}{dr}r\left[r^{-1/2+P}\right]\right\} =$$
$$= -i\frac{a_{st}^2 \hbar^2}{2m} \lim_{r \to 0} r^{2P+1} = 0 \tag{7.17}$$

The last equality follows because $P > 0$ and so, the extra term vanishes. It vanishes also for regular potentials, because for them $P = l + 1/2 > 0$. Therefore the theorem has a form

$$\frac{d\langle \hat{r} \rangle}{dt} = \frac{i}{\hbar} \langle [\hat{H}, \hat{r}] \rangle \qquad (7.18)$$

both for regular as well as singular potentials. As

$$[\hat{H}, \hat{r}] = (-i\hbar) \hat{p}_r / m \qquad (7.19)$$

the final form is

$$\frac{d\langle \hat{r} \rangle}{dt} = \frac{\langle \hat{p}_r \rangle}{m} \qquad (7.20)$$

The obtained results are easily understandable, because the momentum operator is singular at the origin in spite of the coordinate operator.

## 8. Conclusions

In this manuscript we considered influence of the restricted region in 3-dimensional space in the ordinary quantum mechanics, where the radial wave function is defined on a semi-space. Therefore the boundary behavior of radial function contributes to several fundamental relations. The additional contributions appear also from singular behavior of operators under consideration. To our knowledge, *the last fact has not been discussed earlier.*

We have shown that the 3-dimensional consideration involves many significant peculiarities to such problems.

We derived the explicit algorithm of calculation of this extra term and investigated conditions, when it contributes to various fundamental relations.

Application to several known problems shows that the inclusion of the extra term is necessary in order to avoid some misunderstandings.

We believe that the above-developed formalism should have many other applications as well,, especially, in the derivation of uncertainty relations.

## Acknowledgments

First of all, we are grateful to Dr. Ljudmila Gverdciteli, who read our manuscript and gave many useful comments.

This work was supported by Shota Rustaveli National Science Foundation (SRNSF) [grant number № DI-2016-26, Project Title: ''Three-particle problem in a box and in the continuum''].

# APPENDIX .  Comments about the Dirichlet boundary condition

It is known that in spherical coordinates 3-dimensional wave function is represented as

$$\psi(\mathbf{r}) = R(r) Y_l^m(\theta, \varphi) = \frac{u(r)}{r} Y_l^m(\theta, \varphi) \tag{A.1}$$

Correspondingly, after rewritten the Laplacian in terms of polar coordinates two form of radial equations are derived

$$-\frac{\hbar^2}{2m}\left[\frac{d^2}{dr^2} + \frac{2}{r}\frac{d}{dr} - \frac{l(l+1)}{r^2}\right]R(r) + V(r)R(r) = ER(r) \tag{A.2}$$

and

$$-\frac{\hbar^2}{2m}\left[\frac{d^2}{dr^2} - \frac{l(l+1)}{r^2} - V(r)\right]u(r) = Eu(r) \tag{A.3}$$

. P.A.Dirac wrote [34]: "Our equations …strictly speaking are not correct, but the error is restricted by only one point $r = 0$. It is necessary perform a special investigation of solutions of wave equations, that are derived by using the polar coordinates, to be convince are they valid in the point $r = 0$  (p.161)".

Let us discuss briefly the essence of this problem. In the teaching books and scientific articles two methods were applied in the transition from (A.2) to (A.3):
1. The substitution

$$R(r) = \frac{u(r)}{r} \tag{A.4}$$

into Eq. (A.2) or
2. Replacement of the differential expression

$$\left[\frac{d^2}{dr^2} + \frac{2}{r}\frac{d}{dr}\right] \rightarrow \frac{1}{r}\frac{d^2}{dr^2}r \tag{A.5}$$

Now we demonstrate that in both cases the mistakes are made.

After the substitution (A.4) we obtain (only the change in Laplacian is displayed )

$$\frac{1}{r}\left[\frac{d^2}{dr^2} + \frac{2}{r}\frac{d}{dr}\right]u(r) + u(r)\left[\frac{d^2}{dr^2} + \frac{2}{r}\frac{d}{dr}\right]\left(\frac{1}{r}\right) + 2\frac{du}{dr}\frac{d}{dr}\left(\frac{1}{r}\right) \tag{A.6}$$

It is identity. Now the last term cancels the first derivative term in the parenthesis and there remains

$$\frac{1}{r}\frac{d^2u}{dr^2}+u\left[\frac{d^2}{dr^2}+\frac{2}{r}\frac{d}{dr}\right]\left(\frac{1}{r}\right) \tag{A.7}$$

The last term is zero, if we calculate it naively. But really it is a delta function [11]. So we obtain for above expression

$$\frac{1}{r}\frac{d^2u}{dr^2}-4\pi\delta^{(3)}(r) \tag{A.8}$$

Therefore the representation of the Laplacian operator in the form (A.5) is not valid *everywhere*. Results are different in the one point, $r=0$.

If we take into account this fact, we obtain the correct form of equation for the reduced wave function (using polar coordinates also for the delta function)

$$r\left(-\frac{d^2u(r)}{dr^2}+\frac{l(l+1)}{r^2}\right)+\delta(r)u(r)-\frac{2m}{\hbar^2}(E-V(r))ru(r)=0 \tag{A.9}$$

We see that the additional term, containing the delta function, vanishes only when

$$u(0)=0 \tag{A.10}$$

Only in this case we can return to the usual form of reduced equation. Therefore the usual radial equation arises only together with the condition (A.10), which coincides to the Dirichlet boundary condition. No other boundary conditions are permissible for the reduced wave function [34-35].

Therefore, when you use the reduced Schrodinger equation it is necessary to impose the reduced wave function $u(r)$ by the Dirichlet boundary condition (A.10) both for regular as well as singular potentials.

Among the listed papers the 3-dimensional case is considered only in [8].There are two examples for the Coulomb and oscillator potentials, studied by the reduced Schrodinger equation. In addition the Robin boundary condition $u'(0)+\alpha u(0)=0$ is used, which is not correct as follows from above consideration. Here $\alpha$ is a self-adjoint extension parameter. They wrote: "The case of wave functions that vanish at the origin (the standard or the Dirichlet boundary condition for the hydrogen atom) is recovered when $\alpha \to -\infty$ and $u(0)\to 0$, while the product $u(0)\alpha$ remains finite". In $l=0$ for Coulomb potential $V=-\frac{k}{r}$ they write the modified form of virial theorem

$$A-\langle u_n|\frac{\kappa}{r}|u_n\rangle=2E_n \tag{A.11}$$

Here $A$ stands for the extra contribution, $\Pi$ in our notation, they derived (*regularized version*)

$$A_\varepsilon = -\frac{\hbar^2}{2m}|u_n(0)|^2\left(\xi\ln|\xi|\varepsilon + \xi - \alpha + ...\right) \qquad (A.12)$$

In the limit $\alpha \to -\infty$, $u(0) \to 0$, it follows

$$A_\varepsilon = -\frac{\hbar^2}{2m}|u_n(0)|^2\alpha = \frac{\hbar^2}{2m}u_n^*(0)u_n(0)\alpha \qquad (A.13)$$

As $u(0)\alpha$ remains finite and $u(0) \to 0$, one obtains $A_\varepsilon = 0$ and according to (A.11), we return to the usual virial theorem $-\langle u_n|\frac{\kappa}{r}|u_n\rangle = 2E_n$, from (5.4) for $s = l = 0$. Here $k = e^2$.

The same correspondence happens in case of harmonic oscillator.

Therefore, our modified virial theorem with Dirichlet boundary condition for $l = 0$ states gives the same results, as extended radial Hamiltonian with the Robin boundary condition [8]. In our case the procedure of self-adjoint extension is not need.

We have modified the more general hypervirial theorem in the framework of Dirichlet boundary condition, therefore Eqs. (2.16)-(2.17) are new.